\documentclass[aps,pra,floatfix,twocolumn,floatfix,showpacs]{revtex4}

\begin{document}

\title{Calculation of the spectrum of the superheavy element $Z=120$}

\author{ T. H. Dinh, V. A. Dzuba, V. V. Flambaum, and J. S. M. Ginges}

\affiliation{School of Physics, University of New South Wales, Sydney NSW 2052, Australia}

\date{\today}

\begin{abstract}

High-precision calculations of the energy levels of the superheavy
element $Z=120$ are presented. The relativistic Hartree-Fock and 
configuration interaction techniques are employed.
The correlations between core and valence electrons are treated 
by means of the correlation potential method and many-body perturbation theory. 
Similar calculations for barium and radium are used to gauge the accuracy of the  
calculations and to improve the {\it ab initio} results.

\end{abstract}


\maketitle

\section{Introduction}

The study of the superheavy elements is now a popular area of research 
driven by the search for the {\it island of stability} in the region 
$Z$=114 to $Z=126$ where shell closures are predicted (see, e.g., 
\cite{bender}). Elements up to $Z=118$ have been synthesized (see, e.g., 
\cite{hofmann,oganessian}
and evidence for the naturally occurring element $Z=122$ has been reported \cite{marinov}.

Experimental efforts are underway to measure the spectra and chemical properties of the 
superheavy elements \cite{schaedel}. There are also many theoretical works in 
atomic physics and quantum chemistry devoted to these studies (see, e.g., Refs. 
in \cite{pershina,eliav}).

In our previous work \cite{E119} we calculated the electronic spectra of the 
superheavy element $Z=119$ and the singly-ionized superheavy element $Z=120$. 
The nucleus with $Z=120$ protons and $N=172$ neutrons is predicted to be doubly 
magic in relativistic mean-field nuclear calculations \cite{rutz}. 
Moreover, there is evidence from fusion reactions for the enhanced stability of 
the element $Z=120$ \cite{morjean}.

To the best of our knowledge the spectrum of the neutral element $Z=120$ 
has never been calculated before. On the other hand, it has a 
relatively simple electronic structure similar to the structure of its
lighter analogues barium and radium. These atoms have 
two valence electrons above closed shells. The calculations
for the atoms are possible with very high accuracy \cite{DzubaGinges,Ra2}
by combining the correlation potential method with many-body
perturbation theory and the configuration interaction technique. 

In the present work we perform accurate relativistic calculations for the energy 
levels of the neutral superheavy element 120 applying a similar approach as in 
earlier works for barium and radium \cite{DzubaGinges,Ra2}.

\section{Method}

Calculations are performed with a method that 
combines the configuration interaction (CI) technique with many-body
perturbation theory (MBPT). It has been described in detail in our previous 
works \cite{DzubaGinges,Ra2,Kozlov96,vn,vn4}. Here we repeat the main points, 
focusing on the details specific to the current calculations.

Calculations are carried out in the $V^{N-2}$ approximation \cite{vn}. This means 
that the initial Hartree-Fock procedure is performed for the doubly-ionized ion, 
with the two valence electrons removed. This approach has many advantages.
It simplifies the inclusion of the core-valence correlations by avoiding 
the so-called {\it subtraction} diagrams \cite{Kozlov96,vn}. This in turn
allows one to go beyond second order in many-body perturbation theory in 
the treatment of core-valence correlations. Inclusion of the higher-order
core-valence correlations significantly improves the accuracy of the 
results \cite{vn,vn4}. 

The effective CI Hamiltonian for an atom with two valence electrons is the sum
of the two single-electron Hamiltonians and an operator representing the 
interaction between the valence electrons,
\begin{equation}
  H^{\rm eff} = h_1(r_1) + h_1(r_2) + h_2(r_1,r_2) \ .
\label{heff}
\end{equation}
The single-electron Hamiltonian for a valence electron has the form
\begin{equation}
  h_1 = h_0 + \Sigma_1 \ ,
\label{h1}
\end{equation}
where $h_0$ is the relativistic Hartree-Fock Hamiltonian, 
\begin{equation}
  h_0 = c \mbox{\boldmath{$\alpha$}} \cdot {\bf p} + (\beta -1)mc^2 - \frac{Ze^2}{r} + V^{N-2},
\label{h0}
\end{equation}
and $\Sigma_1$ is the correlation potential operator which represents the 
correlation interaction of a valence electron with the core. 

The interaction between valence electrons is given by the sum of the Coulomb interaction
and the correlation correction operator $\Sigma_2$,
\begin{equation}
  h_2 = \frac{e^2}{|\mathbf{r_1 - r_2}|} + \Sigma_2(r_1,r_2) \ .
\label{h2}
\end{equation}
The operator $\Sigma_2$ represents screening of the Coulomb interaction between 
valence electrons by core electrons.

The two-electron wave function for the valence electrons $\Psi$ can be expressed as 
an expansion over single-determinant wave functions,
\begin{equation}
  \Psi = \sum_i c_i \Phi_i(r_1,r_2) \ .
\label{psi}
\end{equation}
The functions $\Phi_i$ are constructed from the single-electron valence basis 
states calculated in the $V^{N-2}$ potential,
\begin{equation}
 \Phi_i(r_1,r_2) = \frac{1}{\sqrt{2}}\big(\psi_a(r_1)\psi_b(r_2)-\psi_b(r_1)\psi_a(r_2)\big) \ .
\label{psiab}
\end{equation}
The coefficients $c_i$ and two-electron energies are found by solving the matrix 
eigenvalue problem
\begin{equation}
  (H^{\rm eff} - E)X = 0 \ ,
\label{Schr}
\end{equation}
where $H^{\rm eff}_{ij} = \langle \Phi_j | H^{\rm eff} | \Phi_i \rangle$ and
$X = \{c_1,c_2, \dots , c_n \}$.

The most complicated part of the calculations is calculation of the
correlation correction operators $\Sigma_1$ and $\Sigma_2$.
We use MBPT and the Feynman diagram technique to do the calculations.
The MBPT expansion for $\Sigma$ starts from the second order in the Coulomb 
interaction. Inclusion of the second-order operators $\Sigma_1^{(2)}$ and $\Sigma_2^{(2)}$
into the effective Hamiltonian (\ref{heff}) accounts for most of the core-valence
correlations. However, further improvement is achieved if higher-order
correlations are included into $\Sigma_1$ and $\Sigma_2$. 

We include higher-orders into $\Sigma_1$ in the same way as
for a single valence electron atom \cite{Dzuba89}. Two dominating
classes of higher-order diagrams are included 
by applying the Feynman diagram technique to the part of $\Sigma_1$
that corresponds to the direct Coulomb interaction. These two classes correspond to 
(a) screening of the Coulomb interaction between valence and core electrons
by other core electrons and (b) the interaction between an electron excited 
from the core and the hole in the core created by this excitation \cite{Dzuba89}

\begin{table}
\caption{Screening factors $f_k$ for inclusion of higher-order 
correlations into the exchange part of $\Sigma_1$ and into
$\Sigma_2$ as functions of the multipolarity $k$ of the Coulomb interaction.}
\label{tb:fk}
\begin{ruledtabular}
\begin{tabular}{ccccccc}
$k$ & 0 & 1 & 2 & 3 & 4 & 5 \\
\hline
$\Sigma^{\rm exch}_1$ & 0.72 & 0.62 & 0.83 & 0.89 & 0.94 & 1.00 \\
$\Sigma_2$            & 0.90 & 0.72 & 0.98 & 1.00 & 1.02 & 1.02 \\ 
\end{tabular}
\end{ruledtabular}
\end{table}

The effect of screening of Coulomb interaction by the core electrons in the 
exchange diagrams is approximated by introducing screening factors $f_k$ (see Table \ref{tb:fk}) 
into each line describing the Coulomb interaction between electrons in a 
Brueckner-Goldstone diagram.
We assume that screening factors $f_k$ depend only on the multipolarity of
the Coulomb interaction $k$. The screening factors were calculated
in our early works \cite{Dzuba88,Dzuba89} and then used in a number of later works.
It turns out that screening factors have very close values for atoms
with similar electron structure. In particular,
the same values can be used for all
atoms of the first and second columns of the periodic table.
The screening factors for $\Sigma^{\rm exch}_1$ were found by 
calculating the direct part of $\Sigma_1$ with and without screening.

A similar way of approximate inclusion of higher-order correlations
via screening factors was used for $\Sigma_2$. The values of the factors, 
however, are different (see Table \ref{tb:fk}).
These factors were found by comparing $\Sigma_1$ in second-order and 
in all-orders with both screening and hole-particle interaction included.

One needs a complete set of single-electron states to calculate $\Sigma$
and to construct the two-electron basis states (\ref{psiab}) for the CI
calculations. We use the same basis in both cases. It is constructed using
the $B$-spline technique \cite{Johnson1,Johnson3}. We use 40 $B$-splines of order 
9 in a cavity of radius $R_{\rm max} = 40 a_B$, where $a_B$ is Bohr's radius.
The upper and lower radial components $R^{u,l}_a(r)$ of the Dirac spinors
for single-electron basis orbitals $\psi_a$ in each partial wave are constructed 
as linear combinations of 40 $B$-splines, 
\begin{equation} 
  R^{u,l}_a(r) = \sum_{i=1}^{40} b^{u,l}_{ai} B_i(r) \ .
\label{bspline}
\end{equation}
The coefficients $b^{u,l}_{ai}$ are found from the condition that $\psi_a$ is an
eigenstate of the Hartree-Fock Hamiltonian $h_0$ (\ref{h0}).

\section{Results}

Our results for barium, radium, and element 120 are presented in Table \ref{tab:results}.
{\it Ab initio} results are listed in the column under the heading ``$\Sigma$''. Here, 
the all-orders correlation potential $\Sigma_1$ is used. This is the same $\Sigma_1$ used in 
our previous work for Ba$^+$, Ra$^+$, and the singly ionized element 120 \cite{E119}. 
For both barium and radium we see that all {\it ab initio} calculated levels lie deeper 
than experiment. That is, the removal energies for the two $s$ electrons from the ground states 
$6s^2$ for barium and $7s^2$ for radium are larger than the experimental results, both by $0.2\%$. 
The {\it ab initio} excitation energies are smaller than experiment. For barium the disagreement 
with measured values is on the order of $1\%$; deviations for the $6s5d$ states are larger 
than for the $6s6p$ states. The results for radium are better.

\begin{table*}
\caption{Energy for removal of two $s$ electrons (for $^1$S$_0$, in atomic units) and excitation 
energies (in units cm$^{-1}$) for barium, radium, and element 120. Results of calculations with all-orders 
correlation potential appear in the column ``$\Sigma$''; those with empirical fitting factors in the 
columns ``$f_{\rm Ba}\Sigma$'' and ``$f_{\rm Ra}\Sigma$''. $g$-factors for element 120 are presented in the 
last column.}
\label{tab:results}
\begin{ruledtabular}
\begin{tabular}{lllcrrrrr}

Atom & Config. & Term & $J$ & $\Sigma$ & $f_{\rm Ba}\Sigma$ & $f_{\rm Ra}\Sigma$ & Exp.\footnotemark[1] & $g$ \\ 
\hline
Barium      & 6s$^2$ & $^1$S & 0 & -0.560223 & -0.559159  & & -0.559159 & \\
            & 6s5d   & $^3$D & 1 & 8687 & 9033 & & 9034 & \\
            & 6s5d   & $^3$D & 2 & 8875 & 9215 & & 9216 & \\
            & 6s5d   & $^3$D & 3 & 9279 & 9601 & & 9597 & \\
            & 6s5d   & $^1$D & 2 & 11081 & 11439 & & 11395 & \\
            & 6s6p   & $^3$P & 0 & 12099 & 12266 & & 12266 & \\
            & 6s6p   & $^3$P & 1 & 12474 & 12634 & & 12637 & \\
            & 6s6p   & $^3$P & 2 & 13365 & 13501 & & 13515 & \\
            & 6s6p   & $^1$P & 1 & 17943 & 18133 & & 18060 & \\
Radium      & 7s$^2$ & $^1$S & 0 & -0.567976 & -0.566819 & -0.566880 & -0.566880 & \\
            & 7s7p   & $^3$P & 0 & 12916 & 13118 & 13088 & 13078 & \\
            & 7s6d   & $^3$D & 1 & 13622 & 13920 & 13719 & 13716 & \\
            & 7s6d   & $^3$D & 2 & 13902 & 14188 & 13993 & 13994 & \\
            & 7s7p   & $^3$P & 1 & 13844 & 14030 & 14001 & 13999 & \\
            & 7s6d   & $^3$D & 3 & 14645 & 14884 & 14712 & 14707 & \\
            & 7s7p   & $^3$P & 2 & 16566 & 16671 & 16652 & 16689 & \\
            & 7s6d   & $^1$D & 2 & 17004 & 17273 & 17094 & 17081 & \\
            & 7s7p   & $^1$P & 1 & 20667 & 20775 & 20739 & 20716 & \\
Element 120 & 8s$^2$ & $^1$S & 0 & -0.626504 & & -0.625256 & & - \\
            & 8s8p   & $^3$P & 0 & 15777 & & 16061 & & - \\
            & 8s8p   & $^3$P & 1 & 17710 & & 17968 & & 1.4214 \\
            & 8s7d   & $^3$D & 1 & 22985 & & 23066 & & 0.5000 \\
            & 8s7d   & $^3$D & 2 & 23163 & & 23231 & & 1.1631 \\
            & 8s7d   & $^3$D & 3 & 23799 & & 23827 & & 1.3332 \\
            & 8s8p   & $^3$P & 2 & 25419 & & 25457 & & 1.4994 \\
            & 8s7d   & $^1$D & 2 & 27438 & & 27477 & & 1.0045 \\
            & 8s8p   & $^1$P & 1 & 27667 & & 27685 & & 1.0766 \\
\end{tabular}
\footnotetext[1]{Ref. \cite{Moore}.}
\end{ruledtabular}
\end{table*}

As a way to empirically correct the results for radium and element 120, we use 
fitting factors found by fitting to measured levels for barium and radium, respectively. 
The empirical fitting factors $f_{\rm Ba}$ and $f_{\rm Ra}$ are different for the different 
partial waves $s$, $p$, and $d$ and placed before the associated $\Sigma_1$. This is done 
in the same way as in our work for single-valence electron systems \cite{E119}. However, 
here the fitting is performed for the two valence electron atom; it means that, as well 
as correcting for unaccounted core-valence correlations, information about the valence-valence 
correlations, incompleteness of the basis, and other higher-order corrections are also contained in this fitting.

Results for spectra calculated with the empirical fitting factors are presented in 
Table \ref{tab:results} under the headings ``$f_{\rm Ba}\Sigma$'' and ``$f_{\rm Ra}\Sigma$''. 
We see that by using the three fitting factors, energy levels for all states considered for 
both barium and radium are in excellent agreement with experiment. This includes the singlet 
states ${^1P}_1$ and ${^1D}_2$. In our previous works on barium and radium \cite{DzubaGinges,Ra2}, 
where similar methods were employed, the singlet states were much worse. This improvement 
comes about from the different screening factors used in $\Sigma_2$ in the current work.

Now we turn to the results for radium and element 120 found with these fitting factors. 
It is seen from Table \ref{tab:results} that with empirical fitting factors from barium, all levels 
for radium move in the direction of experiment. However, the corrections are over-shot.
The triplet $7s7p$ levels are seen to improve, from about -1\% deviation to $\sim 0.1\%$. 
On the other hand, the $7s6d$ levels get worse, with deviations about 1\%. 
We note that the largest deviations from experiment for the {\it ab initio} results for barium are for the 
$6s5d$ levels, with deviations $-$(3-4)\%.
The fitting factor for the $d$-wave has clearly over-compensated these 
corrections.

For element 120, the removal energy for the two $s$ electrons from the ground state $8s^2$ 
decreases by 0.2\% with fitting. 
This is the same difference we saw between {\it ab initio} and measured values for barium and radium. 
For the excitation energies, the differences range from 0.1\% for the higher levels to over 1\% for the 
two lowest levels, with all fitted values larger than the {\it ab initio} ones. We see that the differences 
for element 120 are smaller than for radium. This is because the {\it ab initio} values for radium are 
very good, better than those for barium. We take the fitted results as our final values for the low-lying 
spectra of element 120.

The differences between the {\it ab initio} and fitted values gives an indication of the accuracy of the 
results. While for the higher levels these differences are about 0.1\%, from a consideration of the 
large differences we saw for radium, we will assign a conservative uncertainty of 1\% to our results for 
excitation energies.

We considered Breit and radiative corrections in our work on the spectra of superheavy elements 119 and 120$^+$ 
\cite{E119}. 
We found that by fitting the results of pure correlated calculations to the experimental values for Cs and Ba$^+$, 
the extrapolated values for Fr and Ra$^+$ agreed with experiment at a level better than the estimated size of Breit 
and radiative corrections. This method was then used for the heavier homologues element 119 and 120$^+$ as a way of 
taking into account these corrections semi-empirically. 

In the present work we also fit pure correlated values of the lighter homologues to experiment for use in calculations 
for the heavier elements. In principal, this may be a way of taking into account Breit and radiative corrections. 
However, in this case our calculations are not at the same level of precision as in our work Ref. \cite{E119} 
dealing with one-valence electon atoms and ions. In the current work, the uncertainty in the final 
results stemming from correlations is larger than the expected size of Breit and radiative corrections. 
Therefore, we do not need to account for these corrections.

The ordering of the levels for barium, radium, and element 120 are different. Relativistic 
effects increase approximately as $Z^2\alpha^2$ and are larger for the heavier elements. 
They are responsible for pulling in $s$ and $p$ levels which effectively screen $d$ levels, pushing 
them out. This is the reason that $7s7p \, {^3P}_0$ is the first excited state rather than $7s6d\, {^3D}_1$ 
for radium. For element 120, stronger relativistic effects are responsible for pulling the $8s8p \, {^3P}_1$ 
level below $8s7d \, {^3D}_1$. 

Note that we use the $LS$ notation for the states. 
While the atoms under consideration are highly relativistic and the $jj$ notation is more appropriate, 
the $LS$ notation is used extensively in the literature (including in the tables of Moore \cite{Moore}) 
and for this reason we adopt it here for easier comparison.
In this work we calculated g-factors for element 120 to help in identification of the states; these are presented 
in the final column in Table \ref{tab:results}.

We found in our work Ref. \cite{E119} that there is a sizable volume isotope shift for the elements 119 and 120$^+$. 
In this work, as in Ref. \cite{E119}, calculations have been performed using a two-parameter Fermi distribution 
for the nuclear density with half-density radius $c=8.0$\,fm and 10-90\% width $t=2.0$\,fm corresponding 
to a root-mean-square charge radius $r_{\rm rms}\approx 6.42$\,fm. We define the volume isotope shift as 
\begin{equation}
\frac{\delta E}{E}=k \frac{\delta r_{\rm rms}}{r_{\rm rms}}\, ,
\end{equation}
where here the energies $E$ are taken relative to the continuum (removal energies for two electrons).
Values of $k$ for element 120 are listed in Table \ref{tab:isoshift}. 
A table of values for $r_{\rm rms}$ calculated in the nuclear Hartree-Fock-BCS approximation yields a range 
from around $r_{\rm rms}=6.45$\,fm to $r_{\rm rms}=6.95$\,fm for the light to very heavy isotopes \cite{rms}.
For $r_{\rm rms}=6.90$\,fm, for example, 
the value for the removal energy for the two $s$ electrons in the ground state is $8.5\times 10^{-4}$\,a.u. 
less than for $r_{\rm rms}=6.42$\,fm. This difference amounts to $0.1\%$. 
For the level $8s8p\, {^3P}_0$, the shift is $5.0\times 10^{-4}{\rm \,a.u.}=110{\rm cm}^{-1}$ which is 
$\sim 0.5\%$ of the value for the excitation energy.

\begin{table}
\caption{Volume isotope shifts $k$ for states of element 120.}
\label{tab:isoshift}
\begin{ruledtabular}
\begin{tabular}{cccccc}
State & $8s^2\, {^1}S_0$ & $8s8p\, {^3}P_0$ & $8s8p\, {^3}P_1$ & $8s8p\, {^3}P_2$ & $8s8p\, {^1}P_1$ \\
\hline
  $k$ & -0.018 & -0.012 & -0.012 & -0.012 & -0.013 \\ 
\end{tabular}
\end{ruledtabular}
\end{table}

\section{Conclusion}

The energy levels of low states of the superheavy element $Z=120$ have
been calculated with an uncertainty of about 1\%. These results may be useful 
for experimental studies of this element and for predicting its chemical properties. 

\section*{Acknowledgments}

This work was supported by the Australian Research Council.

\end{document}